\documentclass[12pt]{article}
\usepackage{psfig,epsf}
\usepackage{a4wide,amssymb,graphicx}
\usepackage{cite}

\def\MeV{\textrm{ MeV}}

\newcommand{\beq}{\begin{equation}}
\newcommand{\eeq}[1]{\label{#1} \end{equation}}

\newcommand{\bed}{\begin{displaymath}}
\newcommand{\eed}{\end{displaymath}}
\def\bea{\begin{eqnarray}}
\def\eea{\end{eqnarray}}

\begin{document}

\title{The $\phi$ meson width in the medium from proton induced
$\phi$ production in nuclei}

\maketitle

\begin{center}
{\large{V.K.~Magas, L.~Roca and E.~Oset}}

\vspace{.5cm}
{\it Departamento de F\'{\i}sica Te\'orica and IFIC
 Centro Mixto Universidad de Valencia-CSIC\\
 Institutos de Investigaci\'on de Paterna, Apdo. correos 22085,
 46071, Valencia, Spain}

\end{center}

 \begin{abstract}

 We perform calculations for the production of $\phi$ mesons in
nuclei at energies just above threshold and  study the $A$
dependence of the cross section. We use results for the $\phi$
selfenergy in the medium obtained within a chiral unitary
approach. We find a strong $A$ dependence which is tied to the
distortion of the incident proton and to the  absorption of the 
$\phi$ in its way out of the nucleus. The effect of this latter
process reduces the cross section in about a factor two in heavy
nuclei proving that the $A$ dependence of the cross section bears
valuable information on the $\phi$ width in the nuclear medium.
Calculations are done for energies which are accessible in an
experimental facility like COSY. 

\end{abstract}

\section{Introduction}

 The study of the properties of vector mesons in a nuclear
medium is one of the subjects in hadron physics which receives
continuous attention (see for instance 
Ref.~\cite{Rapp:2000ej}).  Although originally the $\rho$ meson
properties were mostly investigated, with the time the
renormalization of the $\phi$ properties has been taking over as
it is linked to the way  kaons and antikaons are modified in the
nuclear medium, which has also been the subject of intense
study. One of the motivations for this latter work is the
deviation of the $K^-$  selfenergy from the low density theorem,
which is needed to reproduce kaonic  atoms data
\cite{Batty:1997zp,Hirenzaki:2000da,Baca:2000ic,Gal:2001da},
together with  the possibility of formation of kaon condensates
in neutron stars \cite{Kaplan:1986yq}.

    Another reason why the $\phi$ properties have got a renewed
interest is the fact that the medium renormalization in the case
of the $\phi$ is relatively more drastic than that of the
$\rho$. Indeed, predictions of an increase of the $\phi$ width
by a factor five or six \cite{Oset:2001eg,dani} to ten
\cite{Klingl:1998tm}, at normal nuclear matter density,
 have been made using different chiral
approaches.  This large width should in principle be detectable
experimentally and, indeed, different reactions have been
studied or suggested like $\phi$ production in $A-A$ 
collisions \cite{Pal:2002aw,Yokkaichi:wn},
the reaction $\pi^- p \to \phi n$ in nuclei \cite{Klingl:1998tm}
or different methods based on inclusive $\phi$ photoproduction
in nuclei \cite{Oset:2001na,Mosel,daniluis}.

    The width of the $\phi$ has not been the only matter of
concern and the possible shift of the mass has also captured
attention. In that sense,  the $\phi$ mass change has been
studied in several approaches like using effective Lagrangians 
\cite{Klingl:1998tm,Kuwabara:1995ms,Song:1996gw,Bhattacharyya:1997kx,Klingl:1997kf},
QCD sum rules \cite{Asakawa:1994tp,Zschocke:2002mn} or the
Nambu-Jona-Lasinio model  \cite{Blaizot:1991af}. In
ultra-relativistic heavy ion  collisions the $\phi$ width
modification in matter has also been studied in a
dropping meson mass scenario
\cite{Bhattacharyya:1997kx,Ko:tp,Shuryak:1992yy,Lissauer:1991fr,Panda:1993ik,Ko:1994id},
as a result of collisional broadening through $\phi$-baryon
\cite{Smith:1998xu} or $\phi$-meson \cite{Alvarez-Ruso:2002ib}
scattering  processes. All these works point at a sizeable
renormalization of the $\phi$ width and a small mass
shift\footnote{Exceptionally in Ref. \cite{Asakawa:1994tp}  a
mass shift of a few hundred MeV is reported under hot and dense
matter conditions.}. For this reason we will only be concerned
about the $\phi$ width in the medium.

The aim of the present work is to propose a method to determine
the $\phi$ width in the nuclear medium. The traditional method
in the works quoted above (except \cite{daniluis}) is to look
for a broadening of the $\phi$ width reconstructed from the
invariant mass of its decay products. Here, instead, we use a
different philosophy and we investigate the $A$ dependence of
$\phi$ production in $pA$ collisions, in a similar  way as it was
done in \cite{daniluis}  with the $\phi$ photoproduction in
nuclei,  which is the subject of experimental investigation at
Spring8/Osaka \cite{imai}.  The advantage of performing the
reaction slightly above threshold is that one can rule out the
contribution from coherent $\phi$ production which might obscure
the interpretation of the experimental results in \cite{imai}. 
The present reaction, with its particular kinematics, is
amenable of experimental performance at facilities like COSY.


\section{Nuclear effects in the $\phi$ production}

\subsection{General formalism}
\label{s2.1}

In order to implement the relevant nuclear effects in the $\phi$
production cross section we will use a model based on many body
techniques, successfully used in the past in many works
\cite{Salcedo:md,Carrasco:vq} to study the interaction of
different particles with nuclei. The model assumes a local Fermi
sea at each point in the nucleus and provides a very simple and
accurate way to account for the Fermi motion of the initial
nucleon and the Pauli blocking of the final ones. On the other
hand, we have to take into account the distortion of the incoming
nucleon and the final $\phi$ meson in the their way through the
nucleus, which are evaluated in the present work  using an
eikonal approximation. In fact, the $A$ dependence of the effect
due to the final $\phi$ absorption will provide the way to test
the modification of the $\phi$ meson width in nuclear matter, as
we will explain in much more detail in the following, and which
is the main aim of the present work.

Within the local Fermi sea approach the nuclear cross
section can be evaluated, as a first approximation, as:

\begin{equation}
\sigma_A(p_{Lab})=4\int d^3r\int\frac{d^3 k}{(2\pi)^3}
 \Theta(k_F-|\vec{k}|) \sigma_m(p_{Lab},\vec k, \vec r)
\label{eq:sigmaA1}
\end{equation}

\noindent
since $4\int\frac{d^3 k}{(2\pi)^3}\Theta(k_F-|\vec{k}|)=\rho(r)$,
where $k$ is the momentum of the nucleon in the Fermi sea, $k_F$
the Fermi momentum at the local point, $\Theta$ the step
function, $p_{Lab}$ the momentum of the incident proton
and $\sigma_m$ the elementary $pp\to pp\phi$ and 
$pn\to pn\phi$ average cross section in the nuclear medium, which
will be defined later in Eq.~(\ref{eq:sigmam}). 

At this point we can add in
Eq.~(\ref{eq:sigmaA1}) the following eikonal factor to account for
the distortion of the incident proton in its way
through the nucleus till the reaction point:

\begin{equation}
exp\left[-\int_{-\infty}^z\sigma_{pN}(p_{Lab})\rho(\sqrt{b^2+z'^2})dz'\right]
\label{eq:ISI}
\end{equation}

\noindent 
where $z$ and $\vec{b}$ are the position in the beam
axis and the impact parameter, respectively, of the production
point $\vec{r}$ of Eq.~(\ref{eq:sigmaA1}). In Eq.~(\ref{eq:ISI})
$\sigma_{pN}$ is the total $pp$ and $pn$ averaged experimental
cross section, taken from \cite{PDG}, for a given incident
proton momentum.  Eq.~(\ref{eq:ISI}) represents the probability
for the proton to reach the reaction point without having a
collision with the nucleons,  since $\sigma_{pN}\rho$ is the
probability of proton collisions per unit length. 
The use of this eikonal factor will select the one-step processes
and neglect the possibility that the $\phi$ is produced in a
second collision of the nucleon, or a possible excited $\Delta$.
We shall also take these two-step processes into account later on,
but we advance that the $A$ dependence of the $\phi$ production
cross section is already given quite accurately by the one-step
process.

The final $\phi$ absorption in its way out of the nucleus can
also be accounted for by means of a similar eikonal factor and
for the evaluation of the probability of loss of flux per unit
length we can proceed as follows: let $\Pi(p_\phi,\rho(r))$ be
the $\phi$ selfenergy in a nuclear medium as a function of its
momentum, $p_\phi$, and the nuclear density, $\rho(r)$. We have

\begin{equation}
\frac{\Pi}{2\omega_\phi}\equiv V_{\textrm{opt}}
= \ {\cal R}e {V_{\textrm{opt}}} \ + \ i \, 
{\cal I}m {V_{\textrm{opt}}} \ ,
\end{equation}

\noindent
and hence
\begin{equation}
\frac{\Gamma}{2}=- {\cal I}m\frac{\Pi}{2\omega_\phi} \quad ; \qquad
\Gamma=-\frac{{\cal I}m\Pi}{\omega_\phi}
\equiv\frac{d{\cal P}}{dt} \ ,
\end{equation}

\noindent  
where $\omega_\phi$ is the $\phi$ energy and ${\cal P}$ is the $\phi$
decay probability, including nuclear quasielastic and absorption
channels. However, in
what follows we shall only include in ${\cal I}m\Pi$  the absorption channels of the $\phi$, since in
quasielastic $\phi$ collisions the nucleus will be excited but
the $\phi$ will still be there to be observed.
Hence, we have for the probability of loss of
flux per unit length

\begin{equation}
\frac{d{\cal P}}{dl}=\frac{d{\cal P}}{v\,dt}
=\frac{d{\cal P}}{\frac{p_\phi}{\omega_\phi}dt}
=-\frac{{\cal I}m\Pi}{p_\phi} \ .
\end{equation}

\noindent and the corresponding survival probability is given by

\begin{equation}
 exp\left[ -\int_0^{\infty}
dl \frac{(-1)}{p_\phi}
{\cal I}m\Pi(p_\phi,\rho(r'))\right],
\label{eq:FSI}
\end{equation}

\noindent where
 $\vec{r}\,'=\vec{r}+l\frac{\vec p_\phi}{|\vec p_\phi|}$
 with  $\vec{r}$ the $\phi$ production point inside the
nucleus.  The study of the $A$ dependence of the total
nuclear cross section due to the $\phi$ absorption effect,
Eq.~(\ref{eq:FSI}), is the main aim of this work, since it
reflects the modification of the $\phi$ meson width in
nuclear matter.

The study of the $A$ dependence was also done for $\eta$ photoproduction in \cite{ref1}, along the 
same lines as discussed above, in order to determine the inelastic $\eta N$ cross section in a nuclear 
medium. That paper also shows that an alternative BUU approach \cite{cassing} gives similar results 
to the Glauber, or eikonal, approach. 

Another relevant nuclear effect of sizeable consequences is the
binding energy. When studying nuclear reactions with other
beams, for instance photons, one usually neglects the binding
energy, $V_N$, because it affects both the initial and outgoing
nucleons and then it cancels in the $\delta$ function of energy
conservation. Therefore, one usually considers only the kinetic
energy of the nucleons. In the present reaction, with a proton
as incident beam, the same cancellation happens, because we have
two initial and two final nucleons. However,
in order to neglect
$V_N$ of the nucleons we must consider the kinetic energy of the
nucleon beam at the reaction point which is bigger than the
asymptotic value due to the attractive potential felt by the
proton inside the nucleus. This increase in
the kinetic energy can be evaluated considering that
$E_{tot}=E_{kin}(asymptotic)=E_{kin}(local)+V_s(local)$, where
$V_s$ is the potential due to the local Fermi sea, which in the
Fermi model is $V_s(r)=-\epsilon_F(r)=\frac{-k_F^2(r)}{2M}$,
with $k_F=(\frac{3}{2}\pi^2\rho(r))^{1/3}$ and $M$ the nucleon
mass. With these considerations we can define the local initial
proton momentum $\vec p\,'_{Lab}$ such that
 
\begin{equation}
\sqrt{M^2+\vec {p}\,{'}^2_{Lab}}=
\sqrt{M^2+\vec {p}\,^2_{Lab}}+\frac{k_F^2(r)}{2M}.
\label{eq7}
\end{equation}

\noindent
This $p'_{Lab}$ is the incident proton momentum used as input 
to evaluate the elementary cross section.

The elementary cross section in the nuclear medium for
$p(p'_{Lab})+N(k)\to N(p_1)+p(p_2)+\phi(p_\phi)$ reaction is

\begin{eqnarray}\nonumber
\sigma_{m}(p'_{Lab},\vec k,\vec r)&=&
\frac{M}{|\vec p_{Lab}|}\frac{1}{(2\pi)^4}
\int d\Omega_\phi \int dp_\phi p^2_\phi
\int{dp_1}\frac{p_1}{P}\frac{M^2}{2E(p_1)\omega(p_\phi)}\times\\
&\times &
\overline{\sum_{s_i}}\sum_{s_f}|T|^2
\Theta(p_1-k_F(r))\Theta(p_2-k_F(r))\Theta(1-A^2)
\label{eq:sigmam}
\end{eqnarray}

\noindent
where $P=p'_{Lab}+k-p_\phi$, and $A$, providing the cosinus of
the angle between $\vec P$ and $\vec p_1$,
 $A\equiv \frac{1}{2|\vec P||\vec p_1|}
\left\{M^2+\vec
P^2+\vec p\,^2_1-[E(p'_{Lab})+E(k)-E(p_1)-\omega(p_\phi)]^2\right\}$,
with $E(q)=\sqrt{M^2+\vec q\,^2}$, 
$\omega(q)=\sqrt{m_\phi^2+\vec q\,^2}$. In Eq.~(\ref{eq:sigmam}) the azimuthal angle of $\vec{p}_1$ with
respect to $\vec{P}$ has already been integrated, assuming that $|T|^2$ does not depend on this angle. This
is, however, supported by the experiment \cite{Balestra:2000ex} where the angular dependence of 
$pp\to pp\phi$ is almost flat. Hence we assume in what follows $|T|^2$ to be angular independent.

Gathering all these results the final expression for the $\phi$
production cross section in nuclei reads:
$$
\sigma_A(p_{Lab}) = \frac{2}{(2\pi)^7}\frac{M^3}{|\vec{p}_{Lab}|}
\int d^2b\int_{-\infty}^{\infty}dz\, 
exp\left[-\int_{-\infty}^z\sigma_{pN}(p_{Lab})
\rho(\sqrt{b^2+z'^2})dz'\right] 
$$
$$
\int  d^3k \int{dp_1} \int d\Omega_\phi \int dp_\phi
\frac{|\vec{p_\phi}|^2|\vec{p_1}|}{|\vec{P}|E(p_1)\omega(p_\phi)}
\overline{\sum_{s_i}}\sum_{s_f}|T|^2
\, exp\left[ -\int_0^{\infty}
dl \frac{(-1)}{|\vec{p_\phi}|}
{\cal I}m\Pi(p_\phi,\rho(r'))\right]
$$
\begin{equation} \label{eq:sigmaA2}
\Theta(k_F-|\vec{k}|)\Theta(p_1-k_F(r))\Theta(p_2-k_F(r))\Theta(1-A^2) \,. 
\end{equation}

As argued above, the $T$ matrix can be considered a constant for a given energy, which means
that we can divide Eq.~(\ref{eq:sigmaA2}) by the free reaction 
cross section to get rid of $|T|^2$. The free reaction cross
section can be easily evaluated as

\begin{equation}
\sigma_{free}(p_{Lab})=\frac{M^3}{|\vec
p_{Lab}|}\frac{1}{(2\pi)^3}\int dE(p_1)dE(p_2)\Theta(1-B^2)
\Theta(\sqrt{s}-E(p_1)-E(p_2))\overline{\sum_{s_i}}\sum_{s_f}|T|^2
\label{eq:sigmaelfinal}
\end{equation} 

\noindent
where 
$B=(\sqrt{s}-E(p_1)-E(p_2))^2-m_\phi^2-\vec{p_1}^2-\vec{p_2}^2
/(2|\vec{p_1}||\vec{p_2}|)$.

The observable we will evaluate in the results section is
$R\equiv\sigma_A/(A\sigma_{free})$ from
Eqs.~(\ref{eq:sigmaA2}) and (\ref{eq:sigmaelfinal}).

For the evaluation of ${\cal I}m\Pi$ in nuclear matter we use
the results of the model of Ref.~\cite{dani} and its extension to
finite $\phi$-meson momentum done in \cite{daniluis}. This model
is  based on the modification of the $\bar K K$ decay channel in
the medium by means of a careful treatment of the in medium
antikaon selfenergies \cite{Oset:2001eg}. It uses a
selfconsistent coupled channel unitary calculation, based on
effective chiral Lagrangians, and taking into account Pauli
blocking, pion selfenergies and  mean-field potentials of the
baryons (for the S-wave part) and hyperon-hole excitations (for
the P-wave part).

\begin{figure*}[htb]
\begin{center}
\includegraphics[width=9cm,angle=-90]{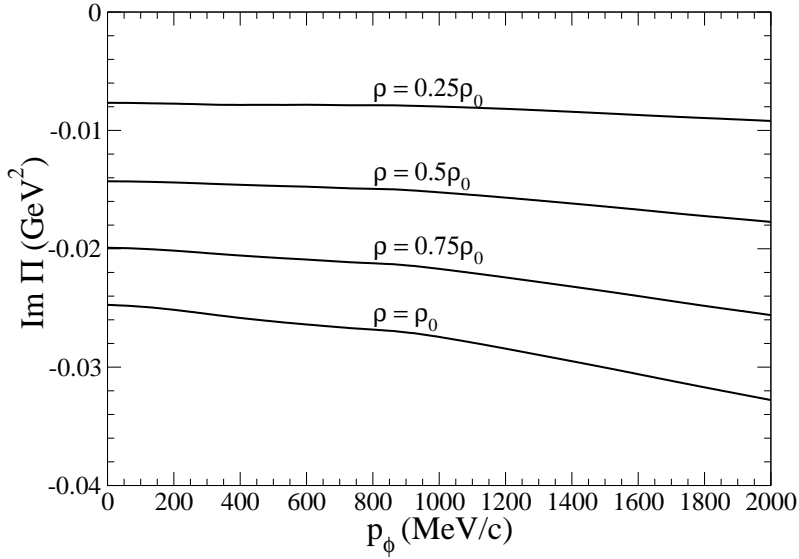}
\end{center}
\caption{Imaginary part of the in medium $\phi$ selfenergy,
without the inclusion of the
free part, as a function of the momentum of the
$\phi$ and the nuclear density. (From Ref.~\cite{daniluis}).
For a $\phi$ at rest and $\rho=\rho_0$ this corresponds to a
$\phi$ in the medium width correction of $24\MeV$. } 
\label{fig:ImPi}
\end{figure*}

In Fig.~\ref{fig:ImPi} we show the imaginary part of the
$\phi$ selfenergy as a
function of the $\phi$ momentum for different nuclear
densities.
The contribution to $\Pi$ coming from the
free $\phi$ decay into $K\bar{K}$, non density dependent,
has been subtracted from the full
$\Pi$ since the $K\bar{K}$ coming from the
free decay would be
detected and counted as a $\phi$ event, hence it does not
contribute to the loss of flux
required in the argument of the exponential in
Eq.~(\ref{eq:FSI}). As shown in \cite{dani}, this corresponds to
a $\phi$ medium width at rest  at $\rho=\rho_0$ of the order of
$24\MeV$.

At this point it is worth mentioning that we have used 
a local (momentum independent) nucleon selfenergy in Eq.~(\ref{eq7}). 
In practice it also has a momentum dependence which is often 
parametrized in terms of an effective mass \cite{mahaux}.
We propose the following expression for the nucleon energy in the medium 
\beq
E(p)=\sqrt{{M^*}^2+\vec{p}\,^2} + M - M^* - \frac{k_F^2}{2M^*}\,,
\eeq{Nse}
which at low momenta gives the  nonrelativistic expression
\beq
E(p)=M + \frac{\vec{p}\,^2}{2M^*}- \frac{k_F^2}{2M^*}
\eeq{nonR}
used in many body approaches \cite{mahaux} with the Thomas Fermi energy 
as the local potential.
Our expression (\ref{Nse}), at the same time, provides a good relativistic 
energy of the nucleon in the low density limit. The effective mass $M^*$ is
evaluated in accurate many body calculations
 \cite{Mstar,Muther:1995bk}, which provide a $\rho$ dependence which can be
approximated by
\beq
M^*=M\left(1-0.2 \rho/\rho_0 \right)\,.
\eeq{Ms}
At high energies, however, the use of Eq.~(\ref{Nse}) with the effective mass of 
Eq.~(\ref{Ms}) would overestimate the effect of nonlocalities, so our calculation 
using this approximation should be considered as an upper bound.

We have evaluated the $\phi$ production cross section modifying Eq.~(\ref{eq7}) to
\beq
\sqrt{{M^*}^2+\vec {p}\,{'}^2_{Lab}} + M - M^* - \frac{k_F^2}{2M^*}=\sqrt{M^2+\vec {p}\,^2_{Lab}}\,,
\eeq{eq7a}
and the expression for A (after Eq.~(\ref{eq:sigmam})) by 
\beq
A\equiv \frac{1}{2|\vec P||\vec p_1|}
\left\{{M^*}^2+\vec
P^2+\vec p\,^2_1-[E(p'_{Lab})+E(k)-E(p_1)-\omega(p_\phi)]^2\right\}\,,
\eeq{eq8a}
with $E(p)=\sqrt{{M^*}^2+\vec{p}\,^2}$. We shall comment in the results section that there are some small
corrections coming from this source, but they do not modify the $A$ dependence of the results.

\subsection{Two-step process with nucleon intermediate state}
\label{s2.2}

Now we want to improve the former result taking into account
$\phi$ production from two-step processes. Let us imagine we
have a $pN$ collision of the initial proton going to any other
channel than $\phi$ production. In such cases the fast incoming
proton will usually survive although with a reduced energy, by
means of which it still can contribute to $\phi$ production. We
estimate the contribution from this mechanism. The first step is
to estimate the energy loss. The total $pN$ cross section for
$T_p=2.5$ to $3$~GeV is around $40\textrm{ mb}$ and consist of
$30$~\% elastic and $70$~\% inelastic, going mostly to several
pion production. Estimating
that the elastic collisions go in average around angles of 
$\theta=30^o$ or bigger in the CM frame since
$\theta\in[0^o,30^o]$ accounts for only $7$~\% of the phase space,
the $p$ energy loss in the lab frame is around $180\textrm{
MeV}$ or more. On the other hand, with an average three pions
produced per collision, the amount of $500\MeV$ energy loss is a
realistic figure. With the percentages given above for the
elastic and inelastic collisions, one comes out with around
$\Delta E \simeq 400\MeV$ loss per collision or more.

The probability that a $\phi$ is produced in a second step, after
a prior $pN$ collision, is easily calculated assuming that the
proton goes still forward after the $pN$ collision, which is
essentially the case in the lab frame.

Since $\sigma_{pN}\rho(\vec b,z'')dz''$ is the probability of a
$pN$ collision in $dz''$ then the formula to evaluate the cross
section for this process is given by Eq.~(\ref{eq:sigmaA2})
substituting

\begin{eqnarray}\nonumber
& & \frac{1}{|\vec p_{Lab}|}\int_{-\infty}^{\infty}dz 
\,exp\left[-\int_{-\infty}^z\sigma_{pN}(p_{Lab})
\rho(\sqrt{b^2+z'^2})dz'\right]  F(p'_{Lab})\\ \nonumber
&\longrightarrow& 
\frac{1}{|\vec p_{Lab}|}\int_{-\infty}^{\infty}dz''
\int_{z''}^{\infty}dz 
\,exp\left[-\int_{-\infty}^{z''}\sigma_{pN}(p_{Lab})
\rho(\sqrt{b^2+z'^2})dz'\right]
\sigma_{pN}(p_{Lab})\rho(\sqrt{b^2+z''^2})\\
&  & \qquad\qquad\qquad\qquad\quad\,\,\,\,
 exp\left[-\int_{z''}^{z}\sigma_{pN}(p_{Lab})
\rho(\sqrt{b^2+z'^2})dz'\right]
F(p''_{Lab})
\label{eq:2steps}
\end{eqnarray}

\noindent where we have introduced a new integration over $z''$,
the point of primary $pN$ collision and the two exponential
factors account for the probability that the proton reaches the
point $z''$ without any  collision, times the probability that
the struck proton reaches the $\phi$ production point without
any other collision. In this sense we guarantee that there is
one and only one primary $pN$ collision. In
Eq.~(\ref{eq:2steps}) the function $F$ corresponds to the rest
of the integral and factors in Eq.~(\ref{eq:sigmaA2}) as a
function of the proton momentum in the nucleus. Hence,
$p''_{Lab}$, which will be the proton momentum after the primary
collision is given by

\begin{equation}
\sqrt{\vec {p}\,{''}^2_{Lab}+M^2}
=\sqrt{\vec {p}\,{'}^2_{Lab}+M^2}-\Delta E .
\end{equation}

\subsection{Two-step processes with $\Delta$ intermediate states}
\label{s2.3}

 In production processes below or close to threshold, multistep processes are
usually important as we have already mentioned. Also it is known that $\Delta$
production in intermediate states is sometimes an efficient way to produce the
final particles. Two methods among many, have 
particularly been used
with success to deal with these multistep processes, and not necessarily around
threshold, but particularly above  it where some of the simplifications done are
more accurate.   One of the methods used is the transport equations
\cite{bertsch,cassing,leupold} (BUU) in which there are collisions of the incoming 
particles which
produce certain final states. In particular, $\Delta$ states are
formed and allowed to propagate till they collide with other
nucleons.   These $\Delta$'s are assumed to be elementary
particles, although the finite lifetime is taken into account. 
Another method which has been used in a large variety of
physical problems is a computer simulation of the reactions
\cite{Salcedo:md,rafaloren,rafamano,amparo},
with similarities to the cascade models. This method,
originally developed to deal with pionic reactions \cite{Salcedo:md},
has also been used to deal with pion \cite{rafaloren} and
nucleon emission \cite{rafamano} in photonuclear reactions and
in nucleon emission in electron nucleus scattering
\cite{amparo}. There is an essential difference between the
transport method and the computer simulation of
\cite{Salcedo:md,rafaloren,rafamano,amparo}.
In the latter one a quantum mechanical many body treatment of the process is done
in which the $\Delta$'s  are never elementary particles, but they appear as
propagators in Feynman many body diagrams, which would qualify as two or three
body processes in the transport method.  The difference is more than technical
since this procedure allows one to sum over the spin of the $\Delta$'s in amplitudes
(in the propagator) while in the transport method the sums over spins are 
done on the cross sections.  This leads sometimes to differences in angular
distributions, the information of which might be missed in the transport
method. However, sometimes, corrections
are done in the standard BUU equations to account for these missing angular
correlations \cite{Muhlich:2004zj}. 
The Monte Carlo simulation 
of \cite{Salcedo:md,rafaloren,rafamano,amparo}
evaluates quantum mechanically these
multistep processes for the diagrams which are called irreducible. In this context reducible diagrams are
those 
that, by cutting a propagator line of a stable particle in the diagram of an amplitude, two valid  
diagrams result, which can be interpreted as a multistep process with stable
particles in the intermediate states.  These processes can be reduced to a sequence of the more
elementary ones. Their probabilities are calculated with the rules for irreducible diagrams, 
and the multistep
ramification is done using a Monte Carlo simulation procedure, allowing the
individual steps to occur according to the calculated probability. 

 Let us address now with the many body techniques,
  using irreducible
 Feynman diagrams, the analogous of the two-step process 
$NN\rightarrow N \Delta$
followed by
$\Delta N\rightarrow NN \phi$ in the transport equation.

\begin{figure*}[htb]
\begin{center}
\includegraphics[height=4cm]{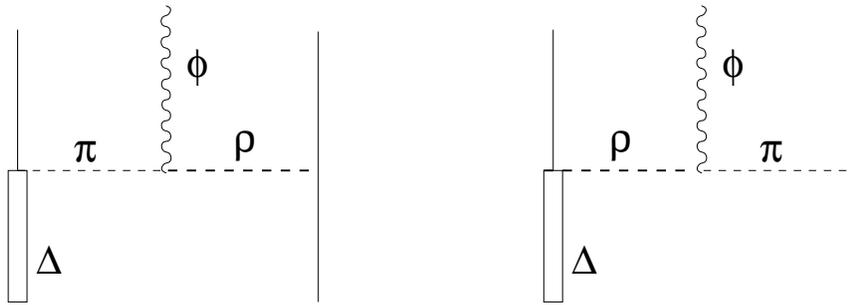}
\end{center}
\caption{Example of one of the possible diagrams for the $\Delta N\rightarrow NN \phi$ reaction 
\cite{titov,barz}.} 
\label{fa1}
\end{figure*}

This two body process would benefit with respect to the one considered in the
former subsection from the fact that 
the $\Delta$  couples more strongly to pions and
vectors than the nucleon. For instance, one can consider a mechanism from the model of 
\cite{titov,barz} like the one on Fig.~\ref{fa1}, which, with respect to the same one with a
nucleon instead of a $\Delta$, would benefit from the factor $f_{\pi N\Delta}/f_{\pi NN}=f_{\rho N\Delta}/f_{\rho NN}=2.13$ 
in the amplitude, hence a factor 4.5 in the cross section.

We shall distinguish between $\Delta$ excitation
on the target and $\Delta$ excitation on the projectile. 
These two
mechanisms appear in the elementary reaction $pp\to\Delta N$ 
\cite{Jain:1986be,Chiba:1991ak,Mundigl:1989af},
but they proved to be relevant in
the $(^3\textrm{He},t)$ reaction in nuclei \cite{Oset:1988cd} and in the
$(\alpha,\alpha')$ reaction. In this latter case,
 only $\Delta$ excitation in the
projectile is allowed \cite{FernandezdeCordoba:1993az}, which was
clearly seen experimentally in \cite{Morsch:1992vj}. We describe
below the two mechanisms starting from the $\Delta$ excitation on
the target.

\begin{figure*}[htb]
\begin{center}
\includegraphics[height=6cm]{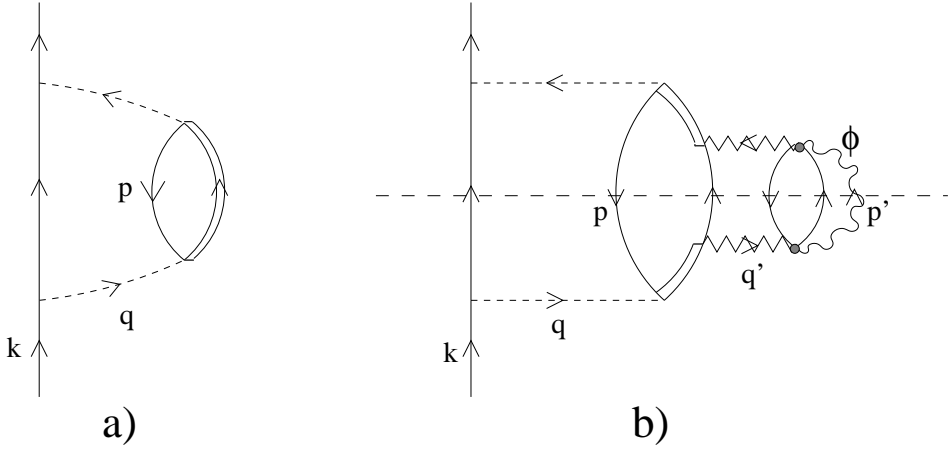}
\end{center}
\caption{Diagram for the $NN\rightarrow N \Delta$ reaction 
for $\Delta$ excitation on the target a). Previous mechanism followed by the
$\Delta N\rightarrow NN \phi$ process b). The long-dashed line represents the cut for the calculation of the
imaginary part of the selfenergy.} 
\label{fa2}
\end{figure*}

We now take the diagram shown in Fig.~\ref{fa2}~b),  which can
be interpreted as having a $NN\rightarrow N \Delta$ collision
followed by $\Delta N\rightarrow NN \phi$. No specific model is
assumed for  $\Delta N\rightarrow NN \phi$,  which is indicated
in Fig.~\ref{fa2}~b) with the serrated line, as we neither did
for the $N N\rightarrow NN \phi$,  but, based on the arguments
given above, we will simple assume that this process has a
cross section about $4.5$ times bigger
 than $NN\rightarrow NN\phi$. 

We evaluate the contribution of the diagram of Fig.~\ref{fa2}~b) to the nucleon selfenergy in the medium. 
A similar diagrammatic approach was used in \cite{barz} to account for
corrections to ordinary multistep
processes but including only nucleons in the intermediate states.  However, it
was rightly concluded there that the equivalent mechanisms with $\Delta$
intermediate states should be more important than the mechanisms evaluated
there.

\begin{figure*}[htb]
\begin{center}
\includegraphics[height=6cm]{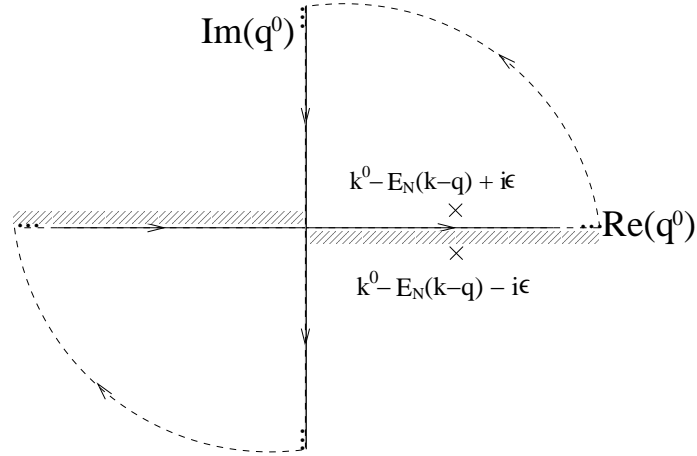}
\end{center}
\caption{Analytic structure and contour used in the $q^0$ integral of Eq.~(\ref{eqa1}) (Wick rotation). The
shadowed regions represent the position of the cuts from the renormalized pion propagator.} 
\label{fa3}
\end{figure*}

We calculate diagram~\ref{fa2}~b)
 in two steps. First we evaluate the
 selfenergy of the incident proton 
 due to the generic diagram of Fig.~\ref{fa2}~a):
\bed
 -i\Sigma(k)=3\int\frac{d^4q}{(2\pi)^4} \left(\frac{f_{\pi NN}} {m_\pi}\right)^2 
 \vec{\sigma}\cdot\vec{q}\ \vec{\sigma}\cdot(-\vec{q})\  
 \frac{i}{{q^0}^2-\vec{q}\,^{2}-m_\pi^2-\Pi_\Delta}\ \frac{ M}{E_N(\vec{k}-\vec{q})}\times
\eed
\beq 
\times \left[\frac{i(1-n(\vec{k}-\vec{q}))}{k^0-q^0-E_N(\vec{k}-\vec{q})+i\epsilon}
+ \frac{i n(\vec{k}-\vec{q})}{k^0-q^0-E_N(\vec{k}-\vec{q})-i\epsilon}\right]\,,
 \eeq{eqa1} 
where the factor $3$ is an isospin factor and $\Pi_\Delta$ represents the pion
selfenergy for $\Delta h$ excitation. One can now perform
a Wick rotation as it is done in \cite{carmen},  
see Fig.~\ref{fa3}.
The integral along the imaginary axis only contributes to the real part of $\Sigma$, and using the contour
in the complex plane of the variable $q^0$ shown in the figure, one picks up only the nucleon
pole of the $1-n(\vec{k}-\vec{q})$ term in Eq.~(\ref{eqa1}),
hence $q^0$ becomes $k^0-E_N(\vec{k}-\vec{q})$. 
One then obtains the imaginary part of the selfenergy, corresponding to the cut shown 
in Fig.~\ref{fa2}~b), as
\beq 
{\cal I}m \Sigma(k) =3 \left(\frac{f_{\pi NN}} {m_\pi}\right)^2 \int\frac{d^3q}{(2\pi)^3} 
(1-n(\vec{k}-\vec{q})){\vec{q}\, }^2
 \frac{{\cal I}m \Pi_\Delta}{\left|(k^0-E_N(\vec{k}-\vec{q}))^2-\vec{q}^{2}-m_\pi^2-\Pi_\Delta
\right|^2}\,,
\eeq{eqa2}    
where ${\cal I}m \Pi_\Delta$ accounts for all possible decay channels of the $\Delta$ 
(like $N\pi$ or $Nph$ or $Nph\phi$).

The mechanism of $\phi$ production will pick up from ${\cal I}m \Pi_\Delta$ in the numerator of 
Eq.~(\ref{eqa2}) only the process with $\phi$ production, but in the denominator ${\cal I}m \Pi_\Delta$ will
appear accounting for all the $\Delta$ decay channels, out of which only the most relevant, $N\pi$,  will be
kept.

$\Pi_\Delta$ can be calculated in the same way and a final
 expression can be found in the appendix of
\cite{pedro}. For our
purpose it is sufficient to take 
\beq
\Pi_\Delta(q)=\frac{4}{9}\left(\frac{f_{\pi N\Delta}} {m_\pi}\right)^2 {\vec{q}\, }^2 \rho \, 
\frac{1}{\sqrt{s_\Delta}-M_\Delta-\frac{M_{\Delta}}{E_{\Delta}(\vec{q})}\Sigma_{\Delta}} 
\,\frac{M_{\Delta}}{E_{\Delta}(\vec{q})}\,,
\eeq{eqa3}
where $\rho$ is the nuclear density and 
$s_\Delta=(q^0+M)^2-{\vec{q}\,}^2$. In Eq.~(\ref{eqa3}) and in
what follows we have neglected the three-momentum, $\vec{p}$, of
the nucleons in the Fermi sea in the energy denominator since
these momenta are small compared to typical values of $q$ and
$k$. Note that we are keeping
 the $M_{\Delta} / E_{\Delta}(\vec{q})$ relativistic factors.
>From Eq.~(\ref{eqa3}) we find 
\beq
{\cal I}m \Pi_\Delta(q)=\frac{4}{9}\left(\frac{f_{\pi N\Delta}}{m_\pi}\right)^2 {\vec{q}\, }^2  \rho 
\left(\frac{M_{\Delta}}{E_{\Delta}(\vec{q})}\right)^2\frac{{\cal I}m \Sigma_{\Delta}}
{\left|\sqrt{s_\Delta}-M_\Delta
-\frac{M_{\Delta}}{E_{\Delta}(\vec{q})}\Sigma_{\Delta}\right|^2}\,.
\eeq{eqa4}

\begin{figure*}[htb]
\begin{center}
\includegraphics[height=4cm]{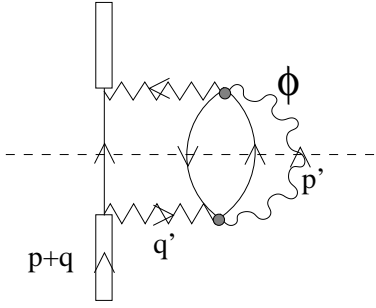}
\end{center}
\caption{$\Delta$ selfenergy diagram accounting
for the $\Delta N\rightarrow NN \phi$. 
The long-dashed line represents the cut for the calculation of the
imaginary part of the selfenergy.} 
\label{fa4}
\end{figure*}

For $\Pi_\Delta$ in the denominator of Eq.~(\ref{eqa2}) it is
sufficient to use Eq.~(\ref{eqa3}) putting
  $\Sigma_{\Delta}= -i \Gamma(s_\Delta)/2$, where
$\Gamma(s_\Delta)$ is the $\Delta$ width for $\pi N$ decay.
However, in the numerator of Eq.~(\ref{eqa2}), 
${\cal I}m \Pi_\Delta$ is given by Eq.~(\ref{eqa4}) where 
${\cal I}m \Sigma_\Delta$ in the numerator should only account
for $\phi$ production. In this way,
when ${\cal I}m \Pi_\Delta$ is placed in the numerator of Eq.~(\ref{eqa2}) it leads to the
relevant term of $\phi$ production shown in Fig.~\ref{fa2}~b). Thus, we need to evaluate the $\Delta$ selfenergy
for the process shown in Fig.~\ref{fa4}.

This diagram can be easily evaluated and gives
(neglecting in the $\Delta$ four-momentum, $p+q$, the
momentum $\vec{p}$ of the nucleon in the Fermi sea):
\bed -i \Sigma_{\Delta}(p+q)
 = \int\frac{d^4q'}{(2\pi)^4} \int\frac{d^4p'}{(2\pi)^4}(-i T) (-i T) 
i U(q'-p')\times 
\eed
\beq
\times \frac{M}{E_N(\vec{q}-\vec{q'})}\ \frac{i(1-n(\vec{q}-\vec{q'}))}
{p^0+q^0-q'^0-E_N(\vec{q}-\vec{q'})+i\epsilon}
\ \frac{i}{{p'^0}^2-{\vec{p'}\, }^2-m_{\phi}^2+i\epsilon}\,,
\eeq{eqa5}
where $U(q'-p')$ is the Lindhard function
 \cite{fetter} with the normalization of the appendix of \cite{pedro}, and we have only kept the particle part
 of the nucleon propagator (the only one which contributes to the imaginary part, as we saw when doing the Wick
 rotation). In Eq.~(\ref{eqa5}) $T$ is the $\Delta N\rightarrow NN\phi$ amplitude, which we simple assume to
 be $2.13$ times bigger than the $N N\rightarrow NN\phi$ one.

 Then ${\cal I}m \Sigma_{\Delta}$ can now be obtained again using a Wick rotation, or easier, 
 applying Cutkosky rules
 \cite{itzykson} adjusted for the present problem \cite{Gil:1997bm} for the cut shown in Fig.~\ref{fa4}: 
\bed
\Sigma_{\Delta} \rightarrow 2i {\cal I}m \Sigma_{\Delta}\,,
\eed
\bed
U(q'-p') \rightarrow 2i \Theta(q'^{0}-p'^{0})
 {\cal I}m U(q'-p')\,,
\eed
\bed
G(p+q-q')\rightarrow 2i \Theta(q^0+p^0-q'^{0}) {\cal I}m G(p+q-q')\,,
\eed
\bed
D(p') \rightarrow 2i \Theta(p'^{0})  {\cal I}m D(p')\,, 
\eed
where $G$ and $D$ are correspondingly nucleon and $\phi$ propagators. 
These rules lead to the expression
\bed
{\cal I}m \Sigma_{\Delta}(k)= \int\frac{d^3q'}{(2\pi)^3}
 \int\frac{d^3p'}{(2\pi)^3}
[1-n(\vec{q}-\vec{q'})][1-n(\vec{q'}-\vec{p'})]|T|^2\times 
\eed
\beq
\times \frac{1}{2\omega_\phi(p')}(-\pi)\, \rho \,
\delta(k^0+2M-E_N(\vec{k}-\vec{q})-E_N(\vec{q}-\vec{q'})-E_N(\vec{q'}-\vec{p'})-\omega_\phi(p')) \,,
\eeq{eqa6} 
which shows explicitly the $\delta$-function
of energy conservation for the process $NNN\rightarrow NNN\phi$.
In Eq.~(\ref{eqa6}) we have used the following approximation:
\begin{equation}
{\cal I}m U(q'-p')\simeq -\pi\rho[1-n(\vec{q}-\vec{q'})]
\delta(q'^{0}-p'^{0}+M-E_N(\vec{q'}-\vec{p'}))
\end{equation}

Altogether we obtain the following expression for ${\cal I}m \Sigma(k)$ in Eq.~(\ref{eqa2}):
\bed
{\cal I}m \Sigma(k)= -\frac{1}{4\pi}\frac{4}{3}\left(\frac{f_{\pi N\Delta}}{m_\pi}\right)^2 
\left(\frac{f_{\pi NN}}{m_\pi}\right)^2 \rho^2 |T|^2
\int\frac{d^3q}{(2\pi)^3} \int\frac{d^3q'}{(2\pi)^3}\int dp'\times 
\eed
\bed
\times \frac{|\vec{p'}|}{|\vec{q'}|}\frac{M}{E_N(\vec{k}-\vec{q})}\frac{M}{E_N(\vec{q}-\vec{q'})} M
{\vec{q}\, }^4
\frac{1}{2\omega_\phi(p')} [1-n(\vec{k}-\vec{q})]
[1-n(\vec{q'}-\vec{p'})][1-n(\vec{q}-\vec{q'})]
\left(\frac{M_{\Delta}}{E_{\Delta}(\vec{q})}\right)^2 \times 
\eed
\bed
\times
\left|\frac{1}{(k^0-E_N(\vec{k}-\vec{q}))^2-\vec{q}^{2}-m_\pi^2-\Pi_\Delta(q)}\right|^2
\left|\frac{1}{\sqrt{s_\Delta}-M_\Delta
-\frac{M_{\Delta}}{E_{\Delta}(\vec{q})}
\Sigma_{\Delta}}\right|^2 F(q)^4\times 
\eed
\beq
\times \Theta(1-\cos(\theta_{\vec{p'}\vec{q'}})^2)\,
\Theta(k^0+2M-E_N(\vec{k}-\vec{q})-E_N(\vec{q}-\vec{q'})-\omega_\phi(p')-M) \,,
\eeq{eqa7} 
with $$\cos(\theta_{\vec{p'}\vec{q'}})=
 \frac{1}{2|\vec p'||\vec q'|}
\left\{M^2+\vec{p'}\,^2+\vec{q'}\,^2-
[k^0+2M-E_N(\vec{k}-\vec{q})-E_N(\vec{q}-\vec{q'})-\omega_\phi(p')]^2\right\}\,.$$ In 
Eq.~(\ref{eqa7}) we have already explicitly introduced a form
factor $F(q)$ for any  $\pi NN$ and $\pi N\Delta$ vertices
\bed
F(q)=\frac{\Lambda^2-m_\pi^2}{\Lambda^2-q^2}
\eed 
with $\Lambda=1\ \textrm{ GeV}$, as usually done.

Next we want to interpret ${\cal I}m \Sigma(k)$ in terms of a cross section for $\phi$ production. 
We follow 
\cite{Marco:1995vb,Gil:1997bm} and write:
\beq
\Gamma=-2\frac{M}{E_N(k)}{\cal I}m \Sigma(k)
\eeq{eqa8}
for the probability of $\phi$ production  per unit time, keeping the important relativistic factors. Then
\beq
\frac{d{\cal P}}{dl}=-2M\frac{{\cal I}m \Sigma(k)}{|\vec{k}|}
\eeq{eqa9}
and the cross section for the process, taking into account the initial proton distortion and 
the final $\phi$ distortion, is given by 
\beq
\sigma=\int d^3r \,\,exp\left[-\int_{-\infty}^z\sigma_{pN}(p_{Lab})
\right]
\left[-2M\frac{{\cal I}m 
\widetilde{\Sigma}(p_{lab})}{|\vec{p}_{lab}|}\right] 
\,,
\eeq{eqa10}
\noindent
where ${\cal I}m \widetilde{\Sigma}(p_{lab})$ is calculated with
the same expression as in Eq.~(\ref{eqa7}) including in addition
the final $\phi$ distortion factor of Eq.~(\ref{eq:FSI}),

\begin{equation}
exp\left[ -\int_0^{\infty} dl \frac{(-1)}{|\vec{p}_\phi|}
{\cal I}m\Pi(p_\phi,\rho(r'))\right].
\end{equation}
\noindent
We simplify the calculations by assuming the $\phi$ to
go in the forward direction, $\vec{p}_{\phi}=(0,0,p')$, which is
a good approximation in the nucleus rest frame, where the
calculations are done, as we have checked.  

\begin{figure*}[htb] \begin{center}
\includegraphics[height=6cm]{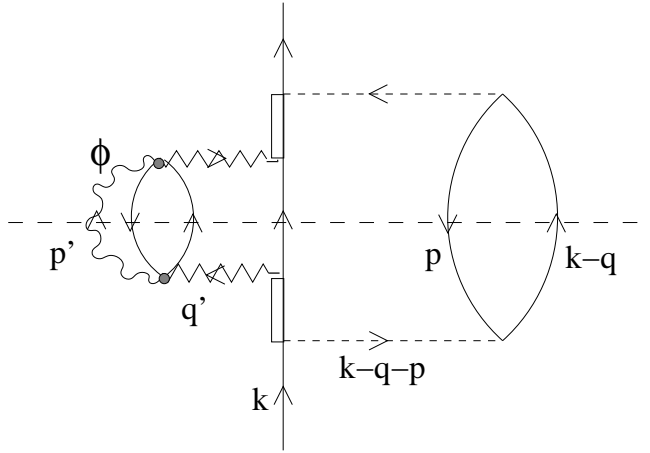} \end{center}
\caption{Diagram for the two-step $\phi$ production with
intermediate $\Delta$ excited on the projectile.} 
\label{fig:deltab} \end{figure*}

Next we briefly describe how to take into account $\Delta$
excitation in the projectile taking advantage of the formalism
described above. The relevant diagram that we should consider is
given in Fig.~\ref{fig:deltab}.

\noindent
The notation for the momenta in Fig.~\ref{fig:deltab} have been
chosen to keep maximum analogy with the previous mechanism.
Hence the same formulae as before can be used adding to  ${\cal
I}m \Sigma(k)$ of Eq.~(\ref{eqa7}) for $\Delta$ excitation in
the target another term  ${\cal I}m \Sigma'(k)$ to account for
$\Delta$ excitation on the projectile, which has the same
expression as Eq.~(\ref{eqa7}) with the following changes:

\begin{eqnarray}
& & {\vec{q}\, }^4 \to (\vec k-\vec q)^4 , \nonumber \\
& &F(q) \to F(k-q-p) ,\nonumber \\
& &\frac{1}{(k^0-E_N(\vec{k}-\vec{q}))^2
-\vec{q}^{2}-m_\pi^2-\Pi_\Delta(q)} \nonumber \\
& &\to \frac{1}{(E_N(\vec{k}-\vec{q})-M)^2-(\vec k-\vec q)^2
-m_\pi^2-\Pi_\Delta(k-q-p)}.
 \label{eq:changes}
\end{eqnarray}

Let us note that the last mechanism is similar to the two-step
mechanism with only  nucleons studied above in subsection
\ref{s2.2}. In the latter case, the projectile nucleon loses
some energy and later on collides with other nucleons to
produce the $\phi$. In the mechanism of Fig.~\ref{fig:deltab}
the projectile gets excited to a $\Delta$ and this $\Delta$
collides with other nucleons to produce the $\phi$. Since we
estimated the cross section for $\Delta N\to NN\phi$ to be
about four times bigger than for $NN\to NN\phi$, we should
expect that the mechanism of Fig.~\ref{fig:deltab} should be
more relevant than the two-step mechanism involving only
nucleons. This is indeed the case, as we shall see in the
results, showing also 
that the mechanism of $\Delta$ excitation in
the projectile is more important than that of $\Delta$
excitation in the target in the present reaction.

It is worth mentioning that this diagrammatic method cannot be
directly used for the two-step nucleon mechanism, because if a
nucleon replaces the $\Delta$ in the diagram of
Fig.~\ref{fig:deltab} this nucleon could be on-shell and would
live forever, thus producing formally an infinite amount of
$\phi$ in its collision with infinite nuclear matter, where the
calculations are done.
 This is not
the case for the $\Delta$ due to the short lifetime.
Hence, in this case the infinite matter approach, together with
the local density approximation (also used in the calculations),
can be reliably used. 
For the nucleon case the explicit
 consideration of the finite size
of the nucleus is essential. This is the reason for the
different treatment of these two processes in spite of their
similarity. These arguments are further elaborated in
\cite{Salcedo:1988xd}.



\section {Results and discussion} 
\label{res}

\begin{figure*}[htb]
\begin{center}
\includegraphics[width=12cm,angle=-90]{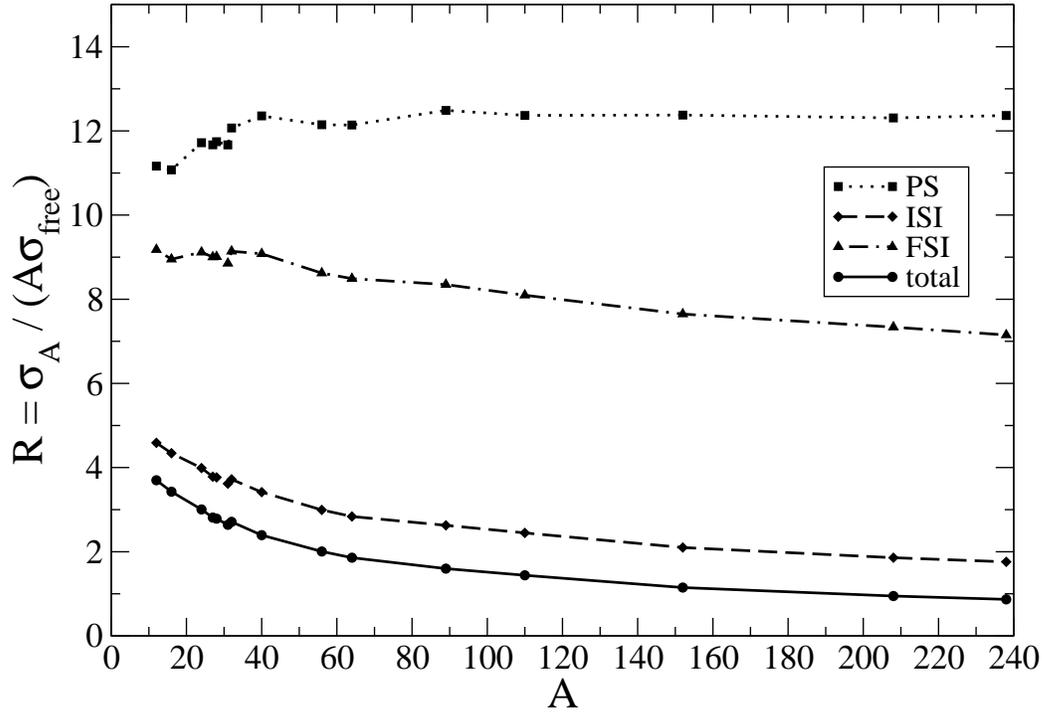}
\end{center}
\caption{The 
results of our calculations for proton kinetic energy
$T_p=2.7\textrm{ GeV}$
with only one-step process.
 Different lines correspond to separate contributions from the
PS, ISI, FSI
and total cross section.} 
\label{ff1}
\end{figure*}

\begin{figure*}[htb]
\begin{center}
\includegraphics[width=12cm,angle=-90]{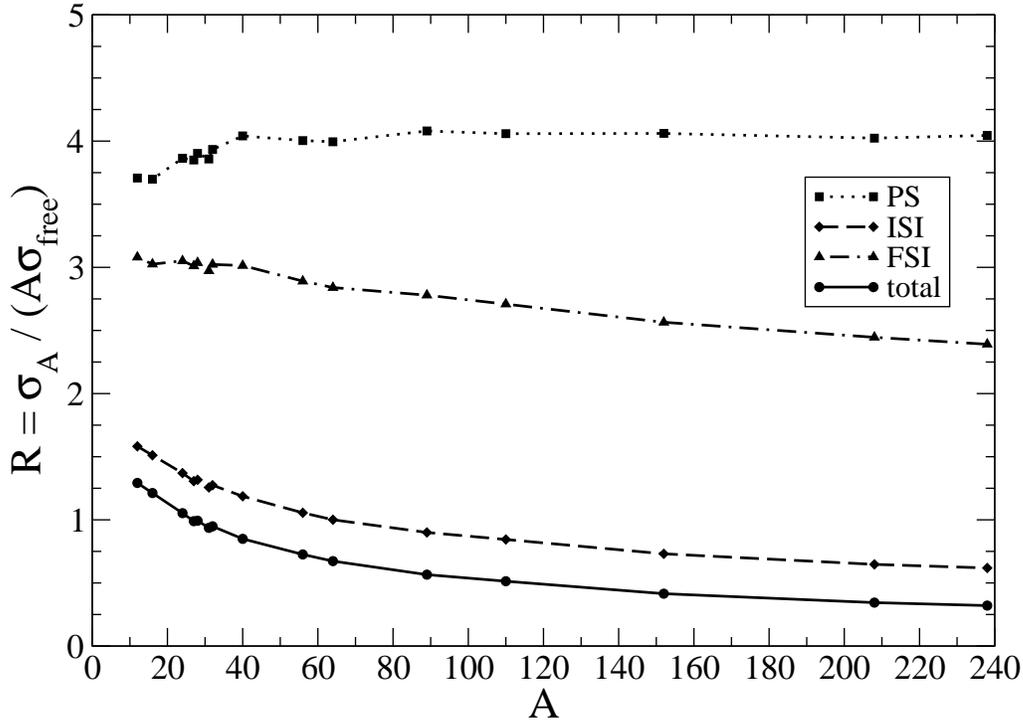}
\end{center}
\caption{The same as Fig.~\ref{ff1}, but for proton kinetic energy energy
$T_p=2.83\textrm{ GeV}$.} 
\label{ff3}
\end{figure*}

First we want to present results from the model
of subsection \ref{s2.1}, accounting for the one-step process.
In Figs.~\ref{ff1} and \ref{ff3}  we show the  results for
$R=\sigma_A/(A\sigma_{free})$ for the projectile energies
$T_p=2.7$ and $2.83 \textrm{ GeV}$ respectively. Different lines
correspond to separate contributions from the phase space (PS),
Eq.~(\ref{eq:sigmaA1}); phase space and initial state
interaction, Eq.~(\ref{eq:sigmaA1}) including the distortion
factor of Eq.~(\ref{eq:ISI}), (ISI);  phase space and final state
interaction, Eq.~(\ref{eq:sigmaA1}) including the distortion
factor of Eq.~(\ref{eq:FSI}), (FSI); and complete calculation,
(total), i.e. the simultaneous contribution of all the effects,
Eq.~(\ref{eq:sigmaA2}). We performed calculations for the
following nuclei:   ${}^{12}_6C$, ${}^{16}_{8}O$, 
${}^{24}_{12}Mg$,  ${}^{27}_{13}Al$, ${}^{28}_{14}Si$, 
${}^{31}_{15}P$,   ${}^{32}_{16}S$,  ${}^{40}_{20}Ca$,  
${}^{56}_{26}Fe$, ${}^{64}_{29}Cu$,  ${}^{89}_{39}Y$,  
${}^{110}_{48}Cd$,  ${}^{152}_{62}Sm$,  ${}^{208}_{82}Pb$,
${}^{238}_{92}U$.

\begin{figure*}[htb]
\begin{center}
\includegraphics[width=11cm,angle=-90]{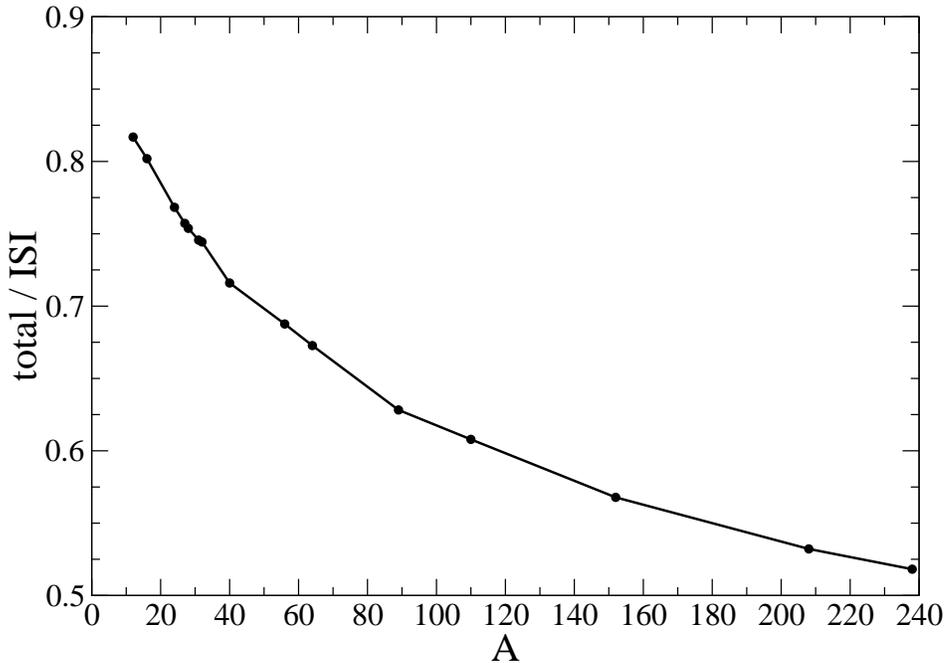}
\end{center}
\caption{Ratio of the total cross section to ISI
with only one-step process. 
 Calculations are done for
$T_p=2.83\textrm{ GeV}$.} 
\label{ff4}
\end{figure*}

 By looking at the PS curves of Figs.~\ref{ff1} and \ref{ff3},
we see that $R$ is larger than unity in both cases and much
larger for the lower $T_p$. This is due to the Fermi motion. 
This increase of the nuclear cross section over $A\sigma_{free}$
is a well known fact in subthreshold production in
particle-nucleus as well as in nucleus-nucleus collisions
\cite{Ko:zp,Barz:2001am}.   The  reaction threshold is
defined for the scattering on a nucleon at rest. Yet, at this
threshold energy, the Fermi motion of the nucleons makes the
reaction possible and the ratio $R$ would grow up to infinity as
we approach the threshold energy. At subthreshold energies, the
consideration of large momentum components of the nucleons in the
nucleus by means of the spectral function, accounting
simultaneously for important correlations between energy and
momentum \cite{Muther:1995bk,deCordoba:1995pt},
 helps to increase the
cross section \cite{Efremov:ua}, as well as the consideration of
multi scattering processes
\cite{Sibirtsev:1997mq,Efremov:1997cq}. All  these mechanisms
become progressively less and less important as we go up to
energies above threshold, and as we discuss below, the possible
uncertainty of our results from neglecting these sources will
not modify our conclusions, since we are  interested in the form
of the $A$-dependence of the cross section rather than in its
absolute value. Indeed, the PS curve is practically constant
with a good accuracy for all $A$. The effects discussed above,
from the spectral function and multinucleon scattering
mechanisms, are volume effects, not affected by the distortion
of the initial proton or the absorption of the final $\phi$.
Thus the constancy in $A$ of the corresponding PS calculations
including these new effects would also hold with a somewhat
increased cross section.  
Now if we look at ISI and FSI curves, we see that in both cases
there is a sizeable decrease of $R$, particularly for ISI, which
shows a stronger $A$ dependence.
Although the ISI and "total" curves are almost parallel,
the absolute values decrease with $A$ and therefore the
contribution of the FSI becomes more and more important.
This significant $A$ dependence can be seen in the ratio of these
two curves which is shown in Fig.~\ref{ff4}.
 We see that the ratio decreases from values around $0.8$
for light nuclei to values  close to $0.5$ for heavy
nuclei. From this figure we can conclude that in the
$A$-dependence there is indeed valuable information concerning
the $\phi$ absorption and hence, the $\phi$ width in the
medium, which is the main conclusion of the present work.

At this point we want to discuss the effect
 of taking into account the 
momentum dependence of the nucleon selfenergy.
 By doing the changes explained in Eqs.~(\ref{eq7a}) and
  (\ref{eq8a}) 
we find an increase of about 15\% in the cross sections (which we would consider as an 
upper bound), but this
percentage increase is remarkably equal for all nuclei such that the A dependence 
of the curves in Figs. \ref{ff1},  \ref{ff3} is preserved. This means that if we normalize the cross 
section to the one of the given nucleus, as we shall do later, the effect discussed 
above does not show up. 

\begin{figure*}[htb]
\begin{center}
\includegraphics[width=12cm]{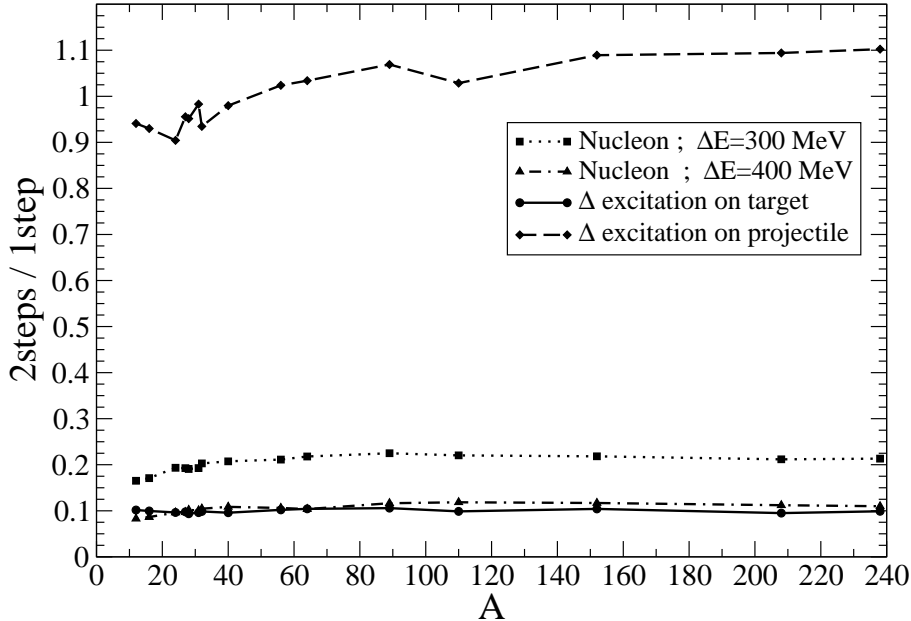}
\end{center}
\caption{Ratio of the nuclear cross section of the different
two-step processes over the cross section of the one-step
process. The
different lines represent: dotted line, only nucleons as
intermediate state with $\Delta E=300\MeV$ (see subsection~\ref{s2.2});
dashed-dotted line, only nucleon with $\Delta E=400\MeV$;
solid line, intermediate $\Delta$ excited on the target (see
Fig.~\ref{fa4}); dashed line,
 intermediate $\Delta$ excited on the projectile (see
Fig.~\ref{fig:deltab}). The calculations are done for a kinetic
energy of the beam proton of $T_p=2.7 \textrm{ GeV}$. } 
\label{ff5a}
\end{figure*}

\begin{figure*}[htb]
\begin{center}
\includegraphics[width=12cm]{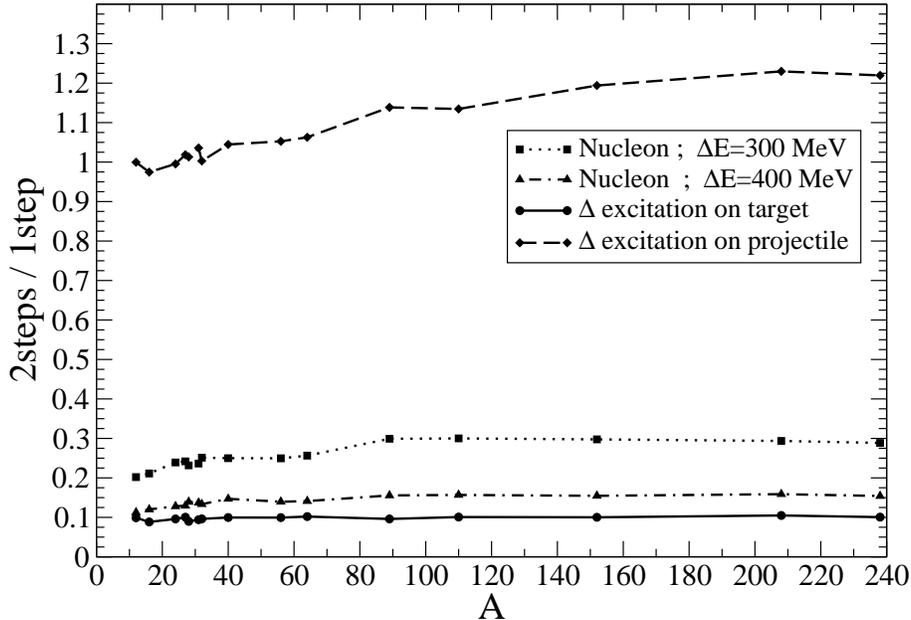}
\end{center}
\caption{Same as Fig.~\ref{ff5a} but for $T_p=2.83 \textrm{ GeV}$.} 
\label{ff5b}
\end{figure*}

Now we pass to consider the contribution from the two-step
processes with nucleon and $\Delta$ intermediate states. 
In Figs.~\ref{ff5a} and \ref{ff5b} we plot the ratios of the
different two-step mechanisms to the one-step process for two
different energies as a function of the mass number.
We see that the most important two-step contribution comes from
the $\Delta$ excitation in the projectile which is comparable to
the one-step mechanism and about $5-10$ times bigger than
the two-step mechanism involving only nucleons or $\Delta$
excitation on the target.
 Even then, we are concerned about the $A$
dependence, no so much on the absolute values of the cross
sections, since the $\phi$ absorption effect is reflected in
this $A$ dependence. 

\begin{figure*}[htb]
\begin{center}
\includegraphics[width=13cm]{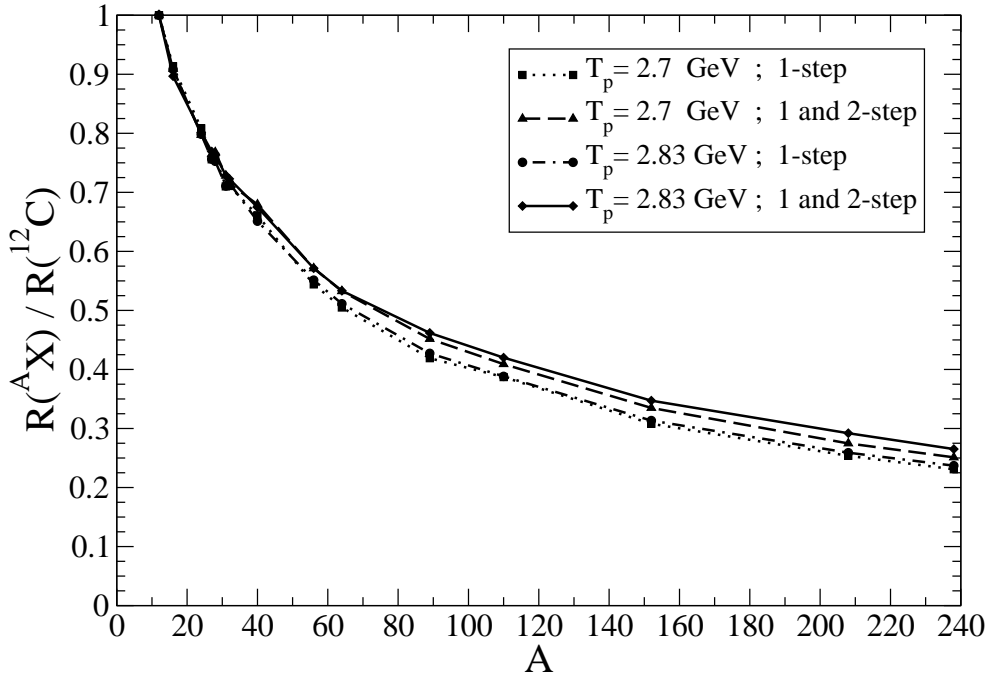}
\end{center}
\caption{Ratio of the nuclear cross section normalized to
$^{12}C$ for two different incident proton kinetic energies,
$T_p$, including or not 
 the two-step mechanisms.
  The two-step process with nucleon intermediate states
 has been evaluated with $\Delta
E=400\MeV$.} 
\label{ff6}
\end{figure*}

The relevance of the two step processes in $\phi$ production was also put of manifest in Ref. \cite{ref2}. There it was shown that the two step mechanism, in which
a pion is produced in the intermediate states was dominant below threshold, but small compared to the one step process at the energies studied here. We estimate our two step cross section involving $\Delta$ in the intermediate states to be of the same order as the one step process. We should note, however, that the two step processes involving $\pi$ or $\Delta$ in the intermediate are not the same, in the sense that the serrated line in Figs. \ref{fa2} and \ref{fig:deltab} does not necessary stand for an on shell pion. Actually, as shown in \cite{barz}, having in this line intermediate $\rho$ states leads to a much larger two step contribution. We accept uncertanties from the two step mechanism, but then we look for one observable that minimizes these uncertanties and this is given by the $A$ dependence of the cross section. 
Indeed,  the fact that 
the curves in Figs.~\ref{ff5a} and \ref{ff5b} are almost
flat as a function of $A$ indicates that the $A$ dependence of
the sum of all mechanisms has essentially the same $A$ dependence
as the one-step mechanism alone. To see this more clearly, 
 we show in Fig.~\ref{ff6}
the ratio $R(^A X)$ normalized to $R(^{12}C)$ for the one-step
and one- plus two-step mechanisms.
 What we see in the figure is that
this normalized $R$ changes very little when including the two-step
mechanisms for both the $T_p$ considered.
Note that for the energy closer to the threshold
($T_p=2.7\textrm{ GeV}$) the changes due to the two-step
contributions are smaller.
 Therefore we conclude that the $A$ dependence obtained
in the present work is  reliable and the calculations
clearly show that proton induced $\phi$ production in nuclei at
energies just above threshold can indeed be used to get
information on the $\phi$ width in the medium.

\begin{figure*}[htb]
\begin{center}
\includegraphics[width=12cm,angle=-90]{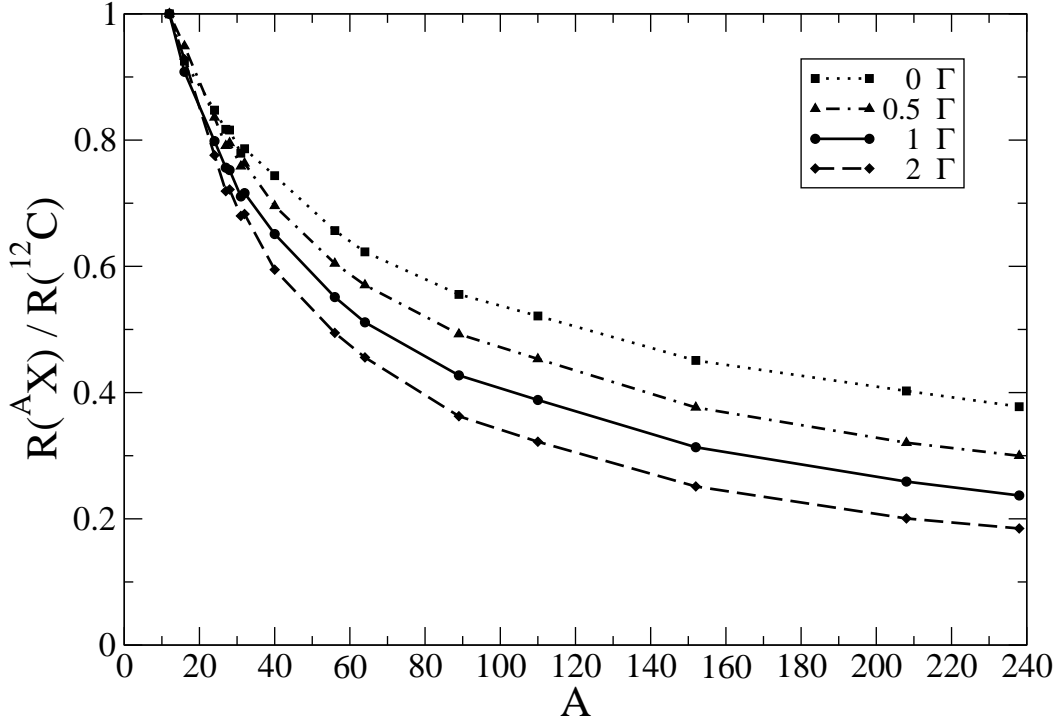}
\end{center}
\caption{Ratio of the nuclear cross section normalized to
$^{12}C$ for $T_p=2.83\textrm{ GeV}$ multiplying the $\phi$ width
in the medium by different factors.} 
\label{ff7}
\end{figure*}

In order to see which is the experimental precision needed to
get a definite information on the $\phi$ width in the medium, we
have performed the same calculations assuming $\phi$ widths in
the medium to be one half or twice the width used so far
\cite{dani,daniluis}.
In the calculations this is implemented by multiplying the
argument of the exponential in Eq.~(\ref{eq:FSI}) by $1/2$ or
$2$ respectively.
 In Fig.~\ref{ff7} we show the results of these
calculations for $T_p=2.83\MeV$ (without the inclusion of the
two-step processes).
Comparing Figs.~\ref{ff6} and \ref{ff7} we clearly see that the
uncertainties due to the second step mechanism are far smaller
than the differences in the results obtained by using these
different $\phi$ widths. The three curves shown there for
the width of \cite{dani,daniluis}, 
half this width, and double, should serve
to get a fair answer about the $\phi$ width in the medium by
comparing with experimental results.
Comparing Figs.~\ref{ff6} and \ref{ff7} one can see
that the uncertainties one might have from the approximate
knowledge of the two-step processes still would allow us to be
sensitive to the value of the $\phi$ width in the medium to the
level of $25$\% of the $\phi$ width we have used.

Now we address another question, which has to do with $\phi$
production in $N\ne Z$ nuclei.
The calculations of the $\phi$ production nuclear cross sections 
are done in symmetric nuclear matter. Hence in
order to calculate the relative  $\phi$
production cross section $R=\sigma_{A}/(A\sigma_{free})$, we
implicitly
took a total free elementary $\phi$ production cross section
$\sigma_{free}=(\sigma_{pn,\phi}+\sigma_{pp,\phi})/2$, 
therefore in a strict sense our model is valid only for the
nuclei with equal amount of $p$ and $n$, i.e. up to
${}^{40}_{20}Ca$ in our calculation. Since the averaged $|T|^2$
used
for elementary $\phi$ production cancels in the numerator
and denominator of $R$, our model can also be
considered for any other series of nuclei with the same ratio of
$Z/N$. Experimentally we have poor knowledge about these
elementary cross sections:  there is experimental data
\cite{Balestra:2000ex} for the  $p\,p\rightarrow p\,p\,\phi$
reaction at one energy, close to the one used in our work, and
nothing for  $p\,n\rightarrow p\,n\,\phi$. Some models (see for
example \cite{titov})  tell us that
$\sigma_{pn,\phi}/\sigma_{pp,\phi}\approx 5$ for our energies.
Nevertheless, our results can still be used to
compare with experiment if
one takes for $\sigma_{free}$ the isospin weighted combination  
$(N\sigma_{pn,\phi}+Z\sigma_{pp,\phi})/A$.
Should we know these elementary 
cross sections we could compare with the
experimental nuclear cross sections
for $N\ne Z$ nuclei multiplying the presents results
for $R$ by $(N\sigma_{pn,\phi}+Z\sigma_{pp,\phi})/A$.
Note, however, that, even with
$\sigma_{pn,\phi}/\sigma_{pp,\phi}=5$, the
ratio 
\begin{equation}
\frac{(N\sigma_{pn,\phi}+Z\sigma_{pp,\phi})/A}
{(\sigma_{pn,\phi}+\sigma_{pp,\phi})/2}\,
\end{equation}
 for a very asymmetric  nucleus like $^{238}U$
is just $1.15$, a small correction compared to the effects from
$\phi$ absorption in this nucleus. However, this correction is
not so small if we consider that the difference in
Fig.~\ref{ff7} for $A\simeq240$ for the full $\phi$ width or twice
this width is only of the order of $33$\% and between the full
width and half this width is of the order of $25$\%. Hence, in
order to determine the medium $\phi$ width with a precision of
better than $50$\% from heavy nuclei the use of this isospin
correction is important.

On the other hand, if we take a nucleus like $^{40}Ca$, which is
isospin symmetric, the ratios of the curves with $2\Gamma$,
$\Gamma$ and $0.5\Gamma$ to the one with $0\Gamma$ are $1.50$,
$1.35$ and  $1.20$ respectively. This gives us an idea of the
precision one can get for $\Gamma_\phi$ given a certain
precision in the experimental results. In our opinion, a way to
get a high precision on this experimental ratio would be to make
a fit to a data set for several approximately symmetric nuclei
and obtain the ratios from the fitted curve.

\section {Conclusions}\label{con} 

We have performed calculations of relative cross sections for
$\phi$ production in nuclei in proton nucleus collisions with
the aim to obtain information on the $\phi$ width in the nuclear
medium. For this purpose we explored the $A$ dependence of the
cross section which is tied to the absorption of the $\phi$ in
its way out of the nucleus.  In the absence of initial proton
and $\phi$ distortions the cross sections obtained are
practically proportional to $A$. Sizeable diversions from this
linear dependence come when both distortions are considered.
Although the initial state interaction was more effective
reducing  the cross section, even then we found sizeable changes
due to $\phi$ absorption which result in an extra reduction of
the cross section in about a factor of two in large nuclei.  
These
predicted changes are large enough  such that devoted
experiments can obtain relevant information on the $\phi$
width in the medium. Since the $A$ dependence of the cross
sections, and not so much the absolute values, is
important to learn about $\phi$ absorption, we present cross
sections for heavy nuclei normalized to a light one, which should
be specially suited for comparison with future experiments.  We
have also seen that in order to extract the optimum information
on the $\phi$ width it would be useful to have data on $\phi$
production on neutron targets, for what experiments on the
deuteron would also be most welcome.    The calculations have
been done at an energy just above threshold, which is accessible
in the COSY facility at Juelich, and for which our theoretical
treatment of the initial state interaction is easy and reliable.
The results obtained in the present work clearly show that the
modification of the $\phi$ width in the nuclear medium has
sizeable effects in this reaction to the point that the actual
experimental implementation of the reaction should
provide a measure of the strength of the medium $\phi$ width.

\section{Acknowledgments}
One of us, L.R., acknowledges support from the
Ministerio de Educaci\'on, Cultura y Deporte.
This work is partly supported by DGICYT contract number BFM2003-00856,
and the E.U. EURIDICE network contract no. HPRN-CT-2002-00311.



\begin{thebibliography}{99}

\bibitem{Rapp:2000ej}
R.~Rapp and J.~Wambach,
Adv.\ Nucl.\ Phys.\  {\bf 25} (2000) 1
[arXiv:hep-ph/9909229].

\bibitem{Batty:1997zp}
C.~J.~Batty, E.~Friedman and A.~Gal,
Phys.\ Rept.\  {\bf 287} (1997) 385.

\bibitem{Hirenzaki:2000da}
S.~Hirenzaki, Y.~Okumura, H.~Toki, E.~Oset and A.~Ramos,
Phys.\ Rev.\ C {\bf 61} (2000) 055205.

\bibitem{Baca:2000ic}
A.~Baca, C.~Garcia-Recio and J.~Nieves,
Nucl.\ Phys.\ A {\bf 673} (2000) 335
[arXiv:nucl-th/0001060].

\bibitem{Gal:2001da}
A.~Gal,
Nucl.\ Phys.\ A {\bf 691} (2001) 268
[arXiv:nucl-th/0101010].

\bibitem{Kaplan:1986yq}
D.~B.~Kaplan and A.~E.~Nelson,
Phys.\ Lett.\ B {\bf 175} (1986) 57.

\bibitem{Oset:2001eg}
E.~Oset and A.~Ramos,
Nucl.\ Phys.\ A {\bf 679} (2001) 616
[arXiv:nucl-th/0005046].

\bibitem{dani}
D.~Cabrera and M.~J.~Vicente Vacas,
Phys.\ Rev.\ C {\bf 67} (2003) 045203
[arXiv:nucl-th/0205075].

\bibitem{Klingl:1998tm}
F.~Klingl, T.~Waas and W.~Weise,
Phys.\ Lett.\ B {\bf 431} (1998) 254
[arXiv:hep-ph/9709210].

\bibitem{Pal:2002aw}
S.~Pal, C.~M.~Ko and Z.~w.~Lin,
Nucl.\ Phys.\ A {\bf 707} (2002) 525
[arXiv:nucl-th/0202086].

\bibitem{Yokkaichi:wn}
S.~Yokkaichi {\it et al.}  [KEK-PS-E325 Collaboration],
Nucl.\ Phys.\ A {\bf 638} (1998) 435.

\bibitem{Oset:2001na}
E.~Oset, M.~J.~Vicente Vacas, H.~Toki and A.~Ramos,
Phys.\ Lett.\ B {\bf 508} (2001) 237
[arXiv:nucl-th/0011019].

\bibitem{Mosel}
P.~Muhlich, T.~Falter, C.~Greiner, J.~Lehr, M.~Post and U.~Mosel,
Phys.\ Rev.\ C {\bf 67} (2003) 024605
[arXiv:nucl-th/0210079].

\bibitem{daniluis}
D.~Cabrera, L.~Roca, E.~Oset, H.~Toki and M.~J.~V.~Vacas,
Nucl.\ Phys.\ A {\bf 733} (2004) 130
[arXiv:nucl-th/0310054].

\bibitem{Kuwabara:1995ms}
H.~Kuwabara and T.~Hatsuda,
Prog.\ Theor.\ Phys.\  {\bf 94} (1995) 1163
[arXiv:nucl-th/9507017].

\bibitem{Song:1996gw}
C.~Song,
Phys.\ Lett.\ B {\bf 388} (1996) 141
[arXiv:hep-ph/9603259].

\bibitem{Bhattacharyya:1997kx}
A.~Bhattacharyya, S.~K.~Ghosh, S.~C.~Phatak and S.~Raha,
Phys.\ Rev.\ C {\bf 55} (1997) 1463
[arXiv:nucl-th/9602042].

\bibitem{Klingl:1997kf}
F.~Klingl, N.~Kaiser and W.~Weise,
Nucl.\ Phys.\ A {\bf 624} (1997) 527
[arXiv:hep-ph/9704398].

\bibitem{Asakawa:1994tp}
M.~Asakawa and C.~M.~Ko,
Nucl.\ Phys.\ A {\bf 572} (1994) 732.

\bibitem{Zschocke:2002mn}
S.~Zschocke, O.~P.~Pavlenko and B.~Kampfer,
Eur.\ Phys.\ J.\ A {\bf 15} (2002) 529
[arXiv:nucl-th/0205057].

\bibitem{Blaizot:1991af}
J.~P.~Blaizot and R.~Mendez Galain,
Phys.\ Lett.\ B {\bf 271} (1991) 32.

\bibitem{Ko:tp}
C.~M.~Ko, P.~Levai, X.~J.~Qiu and C.~T.~Li,
Phys.\ Rev.\ C {\bf 45} (1992) 1400.

\bibitem{Shuryak:1992yy}
E.~V.~Shuryak and V.~Thorsson,
Nucl.\ Phys.\ A {\bf 536} (1992) 739.

\bibitem{Lissauer:1991fr}
D.~Lissauer and E.~V.~Shuryak,
Phys.\ Lett.\ B {\bf 253} (1991) 15.

\bibitem{Panda:1993ik}
A.~R.~Panda and K.~C.~Roy,
Mod.\ Phys.\ Lett.\ A {\bf 8} (1993) 2851.

\bibitem{Ko:1994id}
C.~M.~Ko and D.~Seibert,
Phys.\ Rev.\ C {\bf 49} (1994) 2198
[arXiv:nucl-th/9312010].

\bibitem{Smith:1998xu}
W.~Smith and K.~L.~Haglin,
Phys.\ Rev.\ C {\bf 57} (1998) 1449
[arXiv:nucl-th/9710026].

\bibitem{Alvarez-Ruso:2002ib}
L.~Alvarez-Ruso and V.~Koch,
Phys.\ Rev.\ C {\bf 65} (2002) 054901
[arXiv:nucl-th/0201011].

\bibitem{imai}
K.~Imai, T.~Ishikawa, Private communication.

\bibitem{Salcedo:md}
L.~L.~Salcedo, E.~Oset, M.~J.~Vicente-Vacas and C.~Garcia-Recio,
Nucl.\ Phys.\ A {\bf 484} (1988) 557.

\bibitem{Carrasco:vq}
R.~C.~Carrasco and E.~Oset,
Nucl.\ Phys.\ A {\bf 536} (1992) 445.

\bibitem{PDG}
K.~Hagiwara {\it et al.}  [Particle Data Group Collaboration],
Phys.\ Rev.\ D {\bf 66} (2002) 010001.

\bibitem{ref1}
M.~Effenberger and A.~Sibirtsev,
Nucl.\ Phys.\ A {\bf 632} (1998) 99
[arXiv:nucl-th/9710054].

\bibitem{cassing}
W.~Cassing, V.~Metag, U.~Mosel and K.~Niita,
Phys.\ Rept.\  {\bf 188} (1990) 363.

\bibitem{Balestra:2000ex}
F.~Balestra {\it et al.}  [DISTO Collaboration],
Phys.\ Rev.\ C {\bf 63} (2001) 024004
[arXiv:nucl-ex/0011009].




\bibitem{mahaux}
J.~P.~Jeukenne, A.~Lejeune and C.~Mahaux,
Phys.\ Rept.\  {\bf 25} (1976) 83;
C.~Mahaux {\it et al.},
Phys.\ Rept.\  {\bf 120} (1985) 1.


\bibitem{Mstar}
A. Ramos, A. Polls, W.H. Dickhoff, Nucl.\ Phys.\ A {\bf 503} (1989) 1;
P.~Fernandez de Cordoba and E.~Oset,
Phys.\ Rev.\ C {\bf 46} (1992) 1697;
H.~Muther, A. Polls, W. H. Dickhoff 
Phys.\ Rev.\ C {\bf 51} (1995) 3040.

\bibitem{Muther:1995bk}
H.~Muther, G.~Knehr and A.~Polls,
Phys.\ Rev.\ C {\bf 52} (1995) 2955
[arXiv:nucl-th/9505038].

\bibitem{bertsch}
G.~F.~Bertsch, H.~Kruse and S.~D.~Gupta,
Phys.\ Rev.\ C {\bf 29} (1984) 673.



\bibitem{leupold}
A.~B.~Larionov, M.~Effenberger, S.~Leupold and U.~Mosel,
Phys.\ Rev.\ C {\bf 66} (2002) 054604
[arXiv:nucl-th/0107031].

\bibitem{rafaloren}
R.~C.~Carrasco, E.~Oset and L.~L.~Salcedo,
Nucl.\ Phys.\ A {\bf 541} (1992) 585.

\bibitem{rafamano}
R.~C.~Carrasco, M.~J.~Vicente Vacas and E.~Oset,
Nucl.\ Phys.\ A {\bf 570} (1994) 701.

\bibitem{amparo}
A.~Gil, J.~Nieves and E.~Oset,
Nucl.\ Phys.\ A {\bf 627} (1997) 599
[arXiv:nucl-th/9710070].

\bibitem{Muhlich:2004zj}
P.~Muhlich, L.~Alvarez-Ruso, O.~Buss and U.~Mosel,
Phys.\ Lett.\ B {\bf 595} (2004) 216
[arXiv:nucl-th/0401042].


\bibitem{titov}
A.~I.~Titov, B.~Kampfer and B.~L.~Reznik,
Eur.\ Phys.\ J.\ A {\bf 7} (2000) 543
[arXiv:nucl-th/0001027].

\bibitem{barz}
H.~W.~Barz and B.~Kampfer,
Nucl.\ Phys.\ A {\bf 683} (2001) 594
[arXiv:nucl-th/0005063].

\bibitem{Jain:1986be}
B.~K.~Jain,
Phys.\ Rev.\ C {\bf 32} (1985) 1253.

\bibitem{Chiba:1991ak}
J.~Chiba {\it et al.},
Phys.\ Rev.\ Lett.\  {\bf 67} (1991) 1982.

\bibitem{Mundigl:1989af}
S.~Mundigl and W.~Weise,
Phys.\ Rev.\ C {\bf 39} (1989) 710.

\bibitem{Oset:1988cd}
E.~Oset, E.~Shiino and H.~Toki,
Phys.\ Lett.\ B {\bf 224} (1989) 249.

\bibitem{FernandezdeCordoba:1993az}
P.~Fernandez de Cordoba, Y.~Ratis, E.~Oset, J.~Nieves, M.~J.~Vicente-Vacas, B.~Lopez-Alvaredo and F.~Gareev,
Nucl.\ Phys.\ A {\bf 586} (1995) 586.

\bibitem{Morsch:1992vj}
H.~P.~Morsch {\it et al.},
Phys.\ Rev.\ Lett.\  {\bf 69} (1992) 1336.

\bibitem{carmen}
E.~Oset and L.~L.~Salcedo,
Nucl.\ Phys.\ A {\bf 443} (1985) 704.

\bibitem{pedro}
E.~Oset, P.~Fernandez de Cordoba, L.~L.~Salcedo and R.~Brockmann,
Phys.\ Rept.\  {\bf 188} (1990) 79.

\bibitem{fetter} A.L.~Fetter, J.D.~Walecka, "Quantum theory of many-particle systems", 
New York, McGraw-Hill (1971).

\bibitem{itzykson} C. Itzykson, J.-B. Zuber,   "Quantum field theory", New York, McGraw-Hill (1980).
 
\bibitem{Gil:1997bm}
A.~Gil, J.~Nieves and E.~Oset,
Nucl.\ Phys.\ A {\bf 627} (1997) 543
[arXiv:nucl-th/9711009].

\bibitem{Marco:1995vb}
E.~Marco, E.~Oset and P.~Fernandez de Cordoba,
Nucl.\ Phys.\ A {\bf 611} (1996) 484
[arXiv:nucl-th/9510060].

\bibitem{Salcedo:1988xd}
L.~L.~Salcedo, E.~Oset, D.~Strottman and E.~Hernandez,
Phys.\ Lett.\ B {\bf 208} (1988) 339.



\bibitem{Ko:zp}
C.~m.~Ko,
Phys.\ Lett.\ B {\bf 120} (1983) 294.

\bibitem{Barz:2001am}
H.~W.~Barz, M.~Zetenyi, G.~Wolf and B.~Kampfer,
Nucl.\ Phys.\ A {\bf 705} (2002) 223
[arXiv:nucl-th/0110013].

\bibitem{deCordoba:1995pt}
P.~Fernandez de Cordoba, E.~Marco, H.~Muther, E.~Oset and A.~Faessler,
Nucl.\ Phys.\ A {\bf 611} (1996) 514
[arXiv:nucl-th/9511038].

\bibitem{Efremov:ua}
S.~V.~Efremov and E.~Y.~Paryev,
Phys.\ Atom.\ Nucl.\  {\bf 61} (1998) 541
[Yad.\ Fiz.\  {\bf 61} (1998) 612].

\bibitem{Sibirtsev:1997mq}
A.~Sibirtsev, W.~Cassing, G.~I.~Lykasov and M.~V.~Rzyanin,
Nucl.\ Phys.\ A {\bf 632} (1998) 131
[arXiv:nucl-th/9710044].

\bibitem{Efremov:1997cq}
S.~V.~Efremov and E.~Y.~Parev,
Eur.\ Phys.\ J.\ A {\bf 1} (1998) 99
[arXiv:nucl-th/9701066].

\bibitem{ref2}
 A.~Sibirtsev, W.~Cassing and U.~Mosel,
  Z.\ Phys.\ A {\bf 358} (1997) 357
  [arXiv:nucl-th/9607047].

\end{thebibliography}
\end{document}